\documentclass[nofootinbib,
reprint,
floatfix
]{revtex4-1}

\usepackage{isomath}
\usepackage{upgreek} % to have roman greek letter
\usepackage{siunitx}

\usepackage{epsfig, amsmath,amsfonts, amssymb,graphicx,amsthm,color}
\usepackage{mathtools}% For \MoveEqLeft
\usepackage{dcolumn}% Align table columns on decimal point
\usepackage{bm}% bold math
\usepackage{siunitx}
\usepackage{amssymb}
\usepackage[latin1]{inputenc}
\usepackage{graphicx}
\usepackage{subcaption}
\usepackage{epstopdf}
\usepackage{comment}

\newcolumntype{M}[1]{>{\raggedright}m{#1}}

\begin{document}

\title{Fluorescence and Relaxation Dynamics of Cesium in Argon Matrices: Multiple Trapping Sites and Host-Guest Interactions}
\author{S.  Lahs${{}^{(1)}}^\dagger$}
\author{H.  Dinesan$^{(1)}$}
\author{S.  Mahapatra$^{(1)}$ $^{(4)}$}
\author{W. Chin$^{(2)}$}
\author{C. Crepin$^{(2)}$}
\author{L. Dontot$^{(3)}$}
\author{J. Douady$^{(3)}$}
\author{B. Gervais$^{(3)}$}
\author{D. Comparat$^{(1)}$}
\email[Corresponding author:]{daniel.comparat@cnrs.fr}

\affiliation{$^{(1)}$ Universit\'e Paris-Saclay, CNRS,  Laboratoire Aim\'e Cotton, 91405, Orsay, France. \\ $^{(2)}$ Universit\'e Paris-Saclay, CNRS, Institut des Sciences Mol\'eculaires d'Orsay, 91405, Orsay, France  \\
$^{(3)}$ Laboratoire Cimap, UMR6252 - Universit\'e de Normandie Caen, \'Ecole sup\'erieure
d'ing\'enieures de Caen, Commissariat \`a l'\'energie atomique, Centre national de la recherche
scientifique - 6 Boulevard du Mar\'echal Juin, 14050 Caen Cedex, France \\
$^{(4)}$ Department of Physical Sciences, Indian Institute Of Science Education and Research (IISER) Mohali, 140306, SAS Nagar, Punjab, India.
}
\altaffiliation{Present address: Vienna Center for Quantum Science and Technology, Atominstitut, TU Wien, Vienna, Austria}

\date{\today}

\begin{abstract}

We investigate the fluorescence and relaxation dynamics of Cs atoms embedded in a cryogenic argon matrix using spectroscopy measurements combined with diatomic-in-molecule (DIM) simulations. The data reveal complex emission spectra, large Stokes shifts, and slow relaxation effects, indicating strong host-guest interactions and substantial lattice reorganization. Although the spectra are superimposed on a broad background, possibly due to low-symmetry, defect-related, or grain-boundary trapping sites, the main spectral structure is consistent with two dominant trapping environments that give rise to two triplet absorption features with distinct fluorescence behavior of the doublet and singlet components. Polarization measurements further suggest that these sites may differ in symmetry, although a unique structural assignment remains difficult.
\end{abstract}

\maketitle

\section{Introduction}

 Matrix isolation spectroscopy has been extensively studied since the 1950s \cite{barnes81matrix,almond89,bondybey1996new}. This technique has proven to be a powerful tool for investigating the photophysical and photochemical properties of isolated atoms and molecules embedded in cryogenic rare-gas matrices. By immobilizing reactive or unstable species within a chemically inert environment, it allows precise examination of electronic transitions, relaxation dynamics, and host-guest interactions at low temperatures. Furthermore, the optical transparency and stability of rare-gas matrices make them  suitable for exploring luminescence phenomena of embedded atoms or molecules. 

The interaction between metal atoms and rare-gas hosts has been widely investigated to elucidate weak intermolecular forces, site-specific trapping, and modifications in the electronic structure of the guest species \cite{crepin1999photophysics}. Despite these investigations, it is often difficult to determine how a particular atom is trapped in a specific cryogenic matrix.\\
In the current study, we focus on cesium (Cs) atoms embedded in argon matrix.  Alkali atoms  possess a simple electronic structure, which makes them an ideal model system for describing the trapping of atoms in general. Additionally, Cs-doped matrices are of particular interest for studies of fundamental symmetry violations \cite{weis1997hunting, bouchiat2001atomic, battard2023cesium}. For these studies, a good understanding of the trapping environment is important. In this context, also the possibility of using fluorescence for reading out state-selective populations might be crucial \cite{lancaster2024optical}.

We thus study the trapping environments of Cs atoms in an argon  (Ar) matrix through a combination of absorption, relaxation, and fluorescence spectroscopy, complemented by theoretical simulations. First, we discuss the possible trapping sites. We then investigate the response of the absorption spectrum upon laser irradiation, probing the relaxation mechanisms and the role of disorder (crystallites, defects, ...). Finally, we conduct a detailed luminescence study, using both experimental measurements and wave-packet dynamics simulations, to gain further insight into the excited-state dynamics and lattice relaxation effects.

\begin{figure}
    \centering
    \includegraphics[width=1\linewidth]{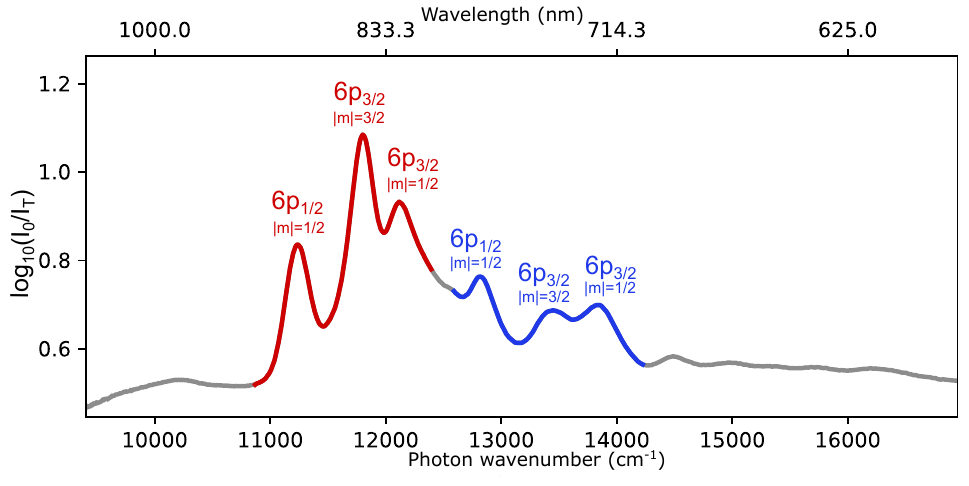}
    \caption{Absorbance spectrum of an Ar crystal doped with Cs at \SI{6}{K}. For labeling simplicity, we choose, for both the red and the blue triplet, to name the bands according to the shift created by a simple Stark shift.}
    \label{fig:CsAbs6K}
\end{figure}

\section{Possible trapping sites of Cs in Ar}

\subsection{Previous experimental results}

In a previous study, we reported on the transmission spectra of Cs doped Ar matrices and their temperature dependency \cite{battard2023cesium}. At temperatures below \SI{16}{K}, we observe six major resonances that can be assigned to two triplet structures (Fig.~\ref{fig:CsAbs6K}). The simplest interpretation was to assign each triplet to a different trapping site. The three lines inside a triplet would then originate from perturbations to the $6s_{1/2}\rightarrow 6p_{1/2}$ and $6s_{1/2}\rightarrow 6p_{3/2}$ transitions.
We label them in analogy to those of a free (gas phase) Cs atom perturbed by an external electric field (see Fig. ~\ref{fig:CsAbs6K}).
%To easily refer to them later on, we choose to label them as done in Fig.~\ref{fig:CsAbs6K}. The label has been chosen according to the shift created by a simple Stark shift, but obviously does not mean that the crystal field is as simple as an electric field on a Cs atom.
There is a relatively large splitting between the components of a given triplet (see Fig.~\ref{fig:CsAbs6K}). The $1/2-3/2$ splitting is significantly larger than the free atom splitting (which is \SI{554}{cm^{-1}}) and the splitting between the two $3/2$ components is also large with respect to the Jahn-Teller splitting observed for lighter alkali in Ar matrices \cite{jacquet2011spectroscopic}.
A stability analysis of the trapping sites in a face-centered cubic (fcc) argon crystal indicated a preference for the T${}_{\rm d}$ (tetrahedral, four vacancies) and the O${}_{\rm h}$ (cubic, six vacancies) symmetries \cite{battard2023cesium}. Absorption spectra revealed satisfactory agreement between the observed and the simulated line positions, although the triplet structure could not fully be reproduced. This could be due to inaccuracies in the close distance part of the excited B $\Upsigma$ state of the CsAr dimer \cite{hewitt2023csar}. Although modifications to its shape might allow for better agreement with the triplet splitting, it seems difficult to achieve this for both trapping sites simultaneously \cite{smai2023etude}. Therefore, the hypothesis of one T${}_{\rm d}$ (4 vacancies) and one O${}_{\rm h}$ (6 vacancies) trapping site is not robust. 

We also observed in previous studies \cite{weyhmann1965optical,Kanda1971} that the relative intensities of the two triplets are difficult to reproduce. We reported that
 the red triplet seems  less stable upon annealing; and a significant background is present (cf. to the offset for the absorbance in Fig. \ref{fig:CsAbs6K}). Also, heating
above \SI{16}{K} leads to permanent changes in the system that can not be undone by cooling the sample down again.
This could suggest the coexistence of other trapping sites, such as a hexagonal close-packed (hcp) Ar component \cite{ozerov2021generic}. This hcp component might also be linked with the fact that solid Ar forms polycrystals at around 10 K with grain sizes around 10 nm \cite{smith1970inert,rudman1978rare,song2013self}. Very similar can be a hexagonal D${}_{3 \rm h}$ (5 vacancies) trapping site in an fcc lattice, such as formed at a
grain boundary resulting from a stacking fault \cite{davis2018investigation}. \\

\subsection{Theory for high symmetry Crystalline trapping sites}

\begin{figure}
    \centering
    \includegraphics[width=1\linewidth, trim=0 0 0 1cm, clip]{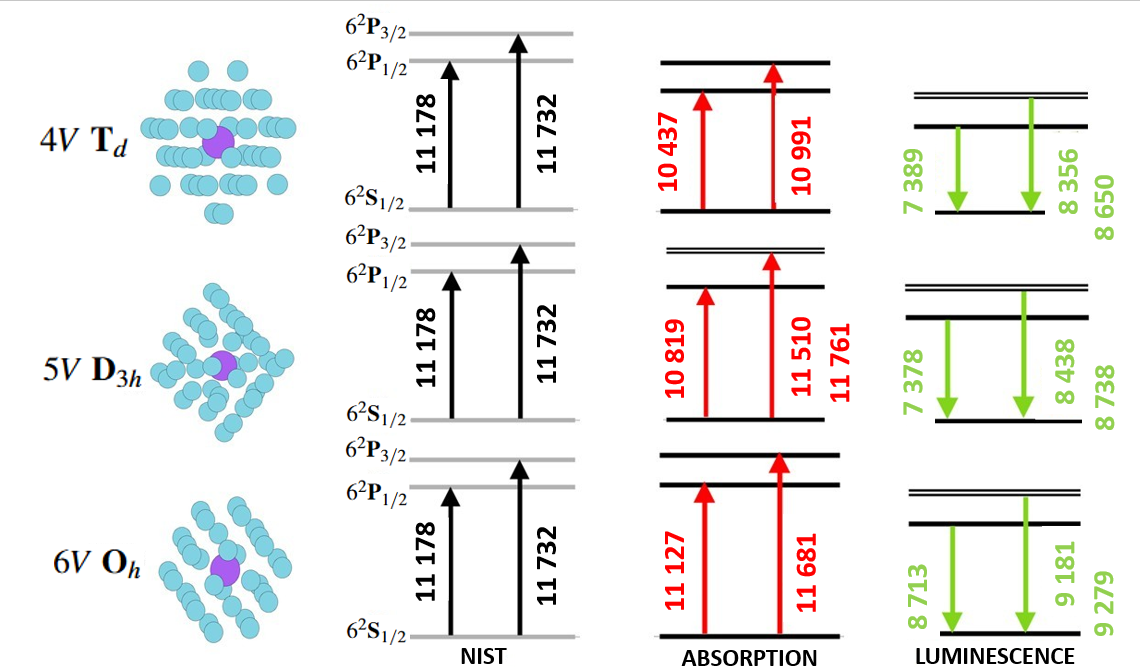}
    \caption{DIM simulation of Cs atoms absorption in the gas phase (black lines \cite{NIST_ASD}), embedded in Ar crystal (red lines)  and emission (green lines after relaxation)  for three high-symmetry trapping sites (T$_{\rm d}$, D$_{\rm {3h}}$ and O$_{\rm h}$). Transition frequencies are in cm$^{-1}$. In left, the sketch Cs is represented as a purple ball surrounded by Ar atoms in blue.}
    \label{fig:DIM_Theory_abs}
\end{figure}

In Fig. \ref{fig:DIM_Theory_abs}, we present the results of a simulation of the trapped Cs transitions
for the two simple fcc O$_{\rm h}$ (six vacancies - 6V) and T$_{\rm d}$ (four vacancies - 4V) but also with a D$_{\rm {3h}}$ (five vacancies - 5V) trapping site.  The simulation relies on simplified pairwise approximation of the atomic interactions using the diatomic-in-molecule (DIM) approach. It considers the Ar ground state and the Cs$(6s)$ and Cs$(6p)$ states, as described in \cite{jacquet2011spectroscopic}. 
The absolute peak location in a DIM simulation has quite a large uncertainty \cite{battard2023cesium}. Therefore, every calculated trapping sites can be seen as compatible with the observed spectra given in Fig. \ref{fig:CsAbs6K}. Due to its asymmetry, the (D$_{\rm {3h}}$ - 5V) trapping site
produced a larger splitting and thus
might provide a more satisfactory explanation of the lifting of the degeneracy of the $J=3/2$ doublet. Whereas in the more symmetric T$_{\rm d}$ and O$_{\rm h}$ sites, it can only be caused by the dynamical Jahn-Teller effect \cite{battard2023cesium}. Simulation of the emission is obtained from the three  Cs(6p) components relaxed in their initial T$_{\rm d}$, D$_{\rm {3h}}$ or O$_{\rm h}$ trapping sites.
For O$_{\rm h}$ and T$_{\rm d}$, the theoretical luminescence spectral lines originate from 
three levels,  including two slightly different components for 6p$_{3/2}$, indicating that this relaxation leads to a reduction in the initial symmetry of the site. 

\subsection{Theory for low symmetry Crystalline trapping sites}

\begin{figure}
    \centering
    \includegraphics[width=1\linewidth, trim=0 0 0 1cm, clip]{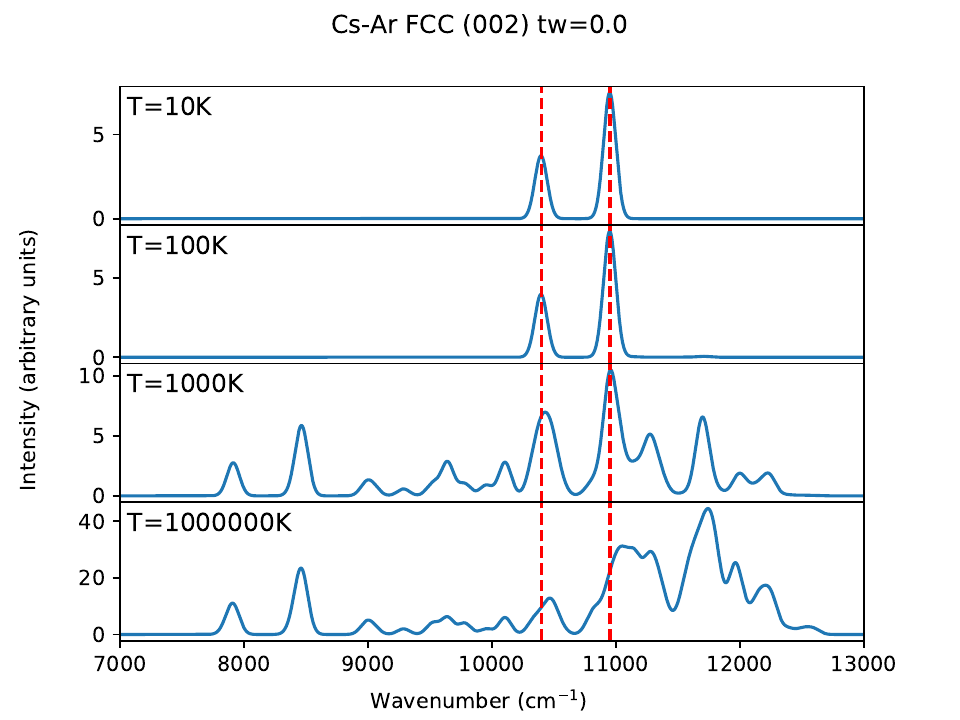}
    \caption{Simulation of the Cs absorption spectrum in an fcc Ar matrix with up to 20 vacancies, as a function of temperature (10 000 Kelvin correspond to a chemical binding energy of roughly 0.86 eV). Each trapping site is weighted by its Boltzmann energy (see text for details). The red lines are a guide for the eyes and refer to the position at $T=0\,$K.}
    \label{fig:fcc_trapping_site}
\end{figure}

We also theoretically investigate larger trapping sites, obtained by removing up to $20$ Ar atoms,  at the center of a large spherical Ar cluster and by relaxing the positions to find the best geometry. Despite their large size, the trapping sites do not collapse during the relaxation process and we obtain a series of well-defined local minima. We use the geometries associated with these minima to produce an absorption spectrum.
Choosing the appropriate weight for a given geometry at a given temperature $T$ is not straightforward, because during the growth of the doped crystal, the process is far from reaching thermal equilibrium. Consequently, metastable states can persist for very long periods, and annealing attempts fail to eliminate them.
%The choice of the weight for a given geometry at a given temperature $T$ is not easy  because during the growth of the doped crystal, the process is far from being fully relaxed towards thermal equilibrium. Therefore metastable state may exist for very long times and attempts of annealing can not get rid of them.
Thus, even if this does not necessarily give any realistic experimental spectrum we make 
 a simple  weight given by the Boltzman distribution $\exp(-\Delta E/k_B T)$, where $\Delta E$ is the estimated energy cost of the trapping site \cite{battard2023cesium}. 
At temperatures below 100 K, only the lowest energy site is populated. On the contrary, for large values of $T$, all sites have a comparable weight. The resulting absorption spectrum shown in Fig. \ref{fig:fcc_trapping_site} provides an illustration of this complexity. The simulated spectrum is well structured, but the observed structure is by no means associated to a well-defined  trapping site.

\subsection{Trapping sites at grain boundaries}

We also investigated trapping sites at grain boundaries.
Indeed,
since the Cs-Ar potential is shallower than Ar-Ar \cite{battard2023cesium} it is difficult to trap Cs atoms in the Ar crystal (similar difficulties have been observed for Na in Ar \cite{ryan2010investigations}), and this may favor the trapping of Cs in defects such as grain boundaries. Furthermore, as mentioned before,  we experimentally observe a large background in the absorption spectrum that could indicate a crystal with numerous defects.

In this theoretical investigation, for simplicity, we only considered the easiest case in which two fcc crystals are rotated against each other while keeping their planes parallel (as illustrated in the left sketch in Fig. \ref{fig:DIM_Theory_abs}).  This type of sites is plausible, since the matrix growth process often results in a sample composed of small grains typically around 10 nanometers in size, and the less compact atomic structure between the grains might be more favorable to accommodate the large Cs atom into the sample. Although we do not know the occurrence of the various grain boundary orientations for Ar crystals, simulations for aluminum obtained by relaxation from an initial random distribution of the atoms suggest that a great number of grain orientations can be obtained \cite{pan1994csl}. Here, we keep the grain boundaries obtained by a twist along the 111 direction by an angle $\theta = 0$ (the same case as for a pure fcc) and $\theta = 60, 38.21, 27.80$ and $46.83$ degrees, labeled respectively by S1, S3, S7, S13 and S19  \cite{pan1994csl}. The corresponding number of independent site orientations for each grain is 3, 16, 21, 30 and 34, respectively. Again, we remove up to 20 Ar atoms to create a cavity associated to each site. All together, we obtain 2144 non-equivalent structures for Cs in Ar. 
The best (lowest energy) trapping site corresponds to a two-atom vacancy for the S7 grain boundary. 
Another plausible trapping site is the  D$_{\rm {3h}}$  five-atom vacancy we have mentioned earlier, corresponding to an S3 grain boundary. 

\begin{figure}
    \centering
    \includegraphics[width=1\linewidth, trim=0 0 0 1cm, clip]{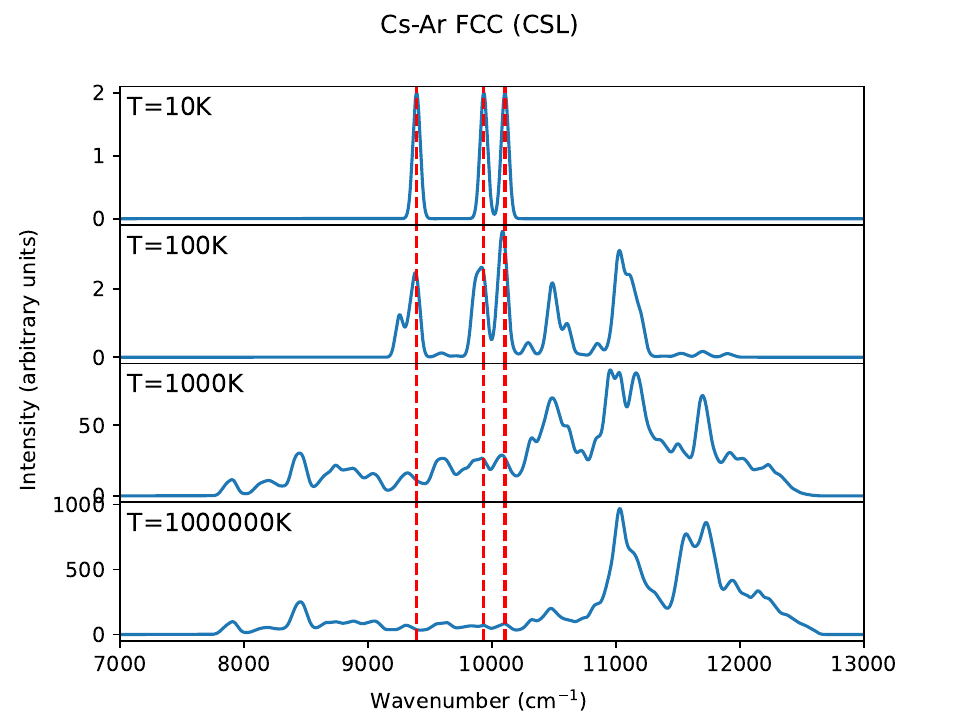}
    \caption{Simulation of the Cs absorption spectrum with various grain boundary orientations for Ar with up to 20 vacancies. Each trapping site is weighted by its Boltzmann energy (see text for detail). Red lines are guide for the eyes and refer to the position at $T=0\,$K.}
    \label{fig:twist}
\end{figure}

 We tentatively apply the same statistical weighting as for the crystalline trapping sites to get a statistical average. The corresponding spectrum is depicted in Fig. \ref{fig:twist}. As for the crystalline sites, the spectrum results from the complex superposition of all the spectra associated to each individual site, but we can no longer associate a series of 2 or 3 peaks to a specific site. At very high temperatures, the statistical averaging effect is comparable to the crystalline site averaging, though the spectrum looks different (but interestingly enough, not too far from the experimental one) because the lowest energy trapping 
site is different. 
 Our theoretical study does not claim to cover all possible trapping scenarios. Nevertheless, it illustrates the potential complexity of the absorption spectrum of Cs in a polycrystalline Ar matrix.

\section{Relaxation process}%of the transmission spectrum upon laser irradiation}

\begin{figure}[t!]
    \centering
    \begin{subfigure}[b]{0.47\textwidth}
        \centering
        \includegraphics[width=\textwidth]{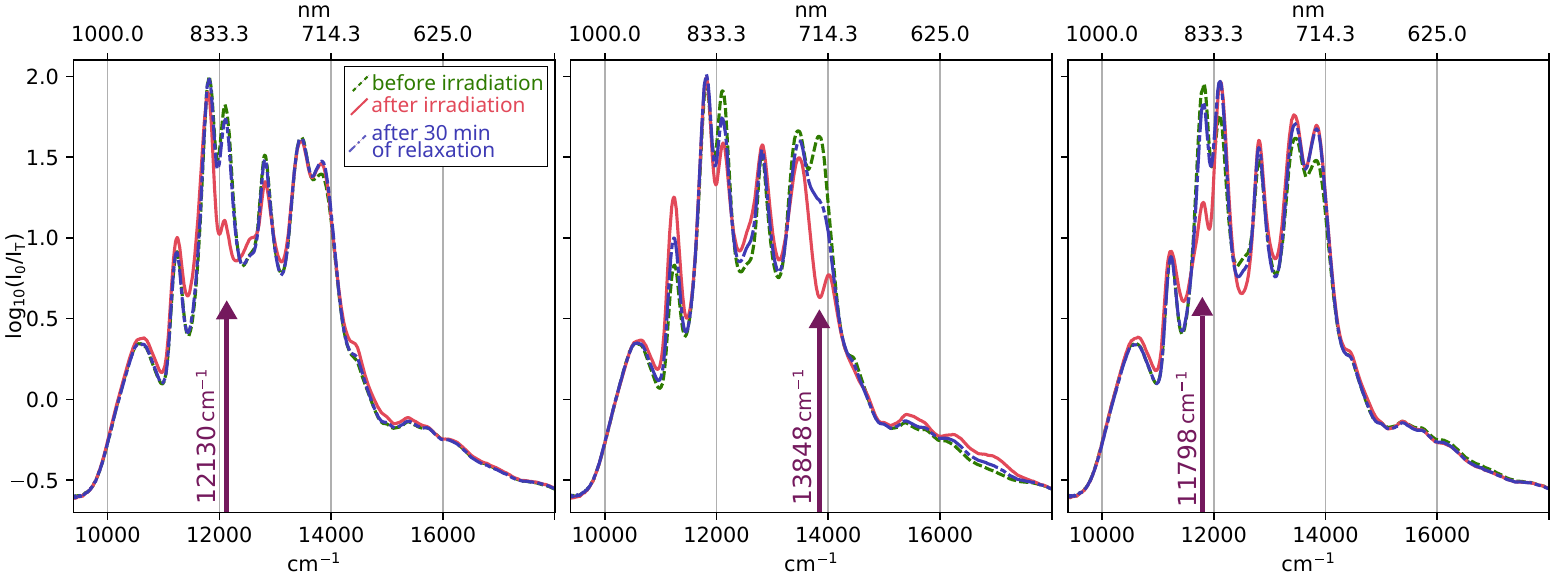}
        \caption{}
        \label{fig:Relaxation}
    \end{subfigure}
    \begin{subfigure}[b]{0.236\textwidth}
        \centering
        \includegraphics[width=\textwidth]{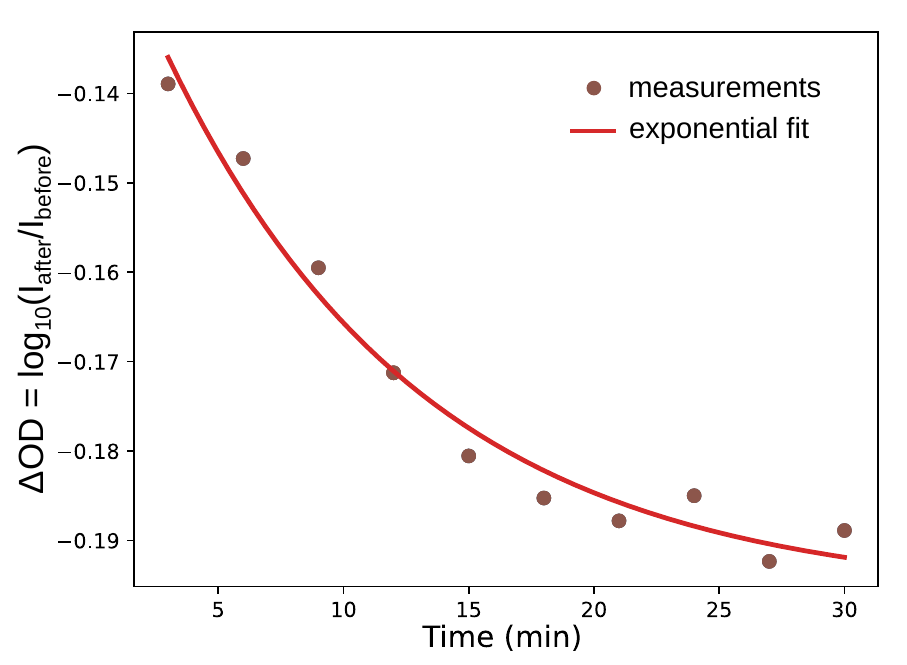}
        \caption{}
        \label{fig:evdecay}
    \end{subfigure}
    \hfill
    \begin{subfigure}[b]{0.236\textwidth}
        \centering
        \includegraphics[width=\textwidth]{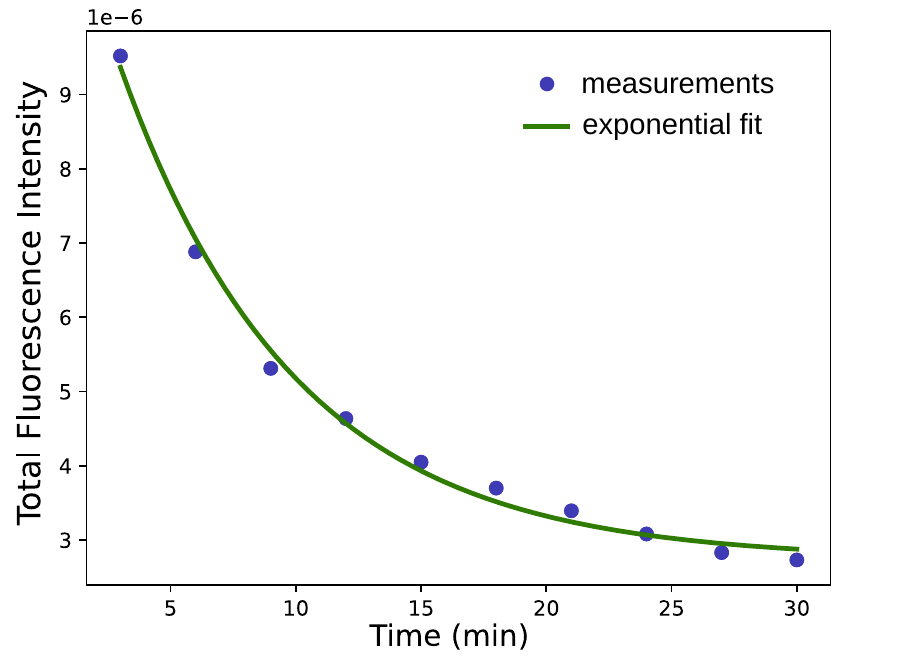}
        \caption{}
        \label{fig:flrdecay}
    \end{subfigure}
    \caption{Evolution of the  absorption spectra under irradiation. Top panel (a): "Relaxation" spectra recorded after laser exposure at three different laser wavelengths (optical depth: $\log_{10}(I_0/I_\text{T})$ along y-axis and wavenumber in cm$^{-1}$ along x-axis). The vertical arrow indicates the excitation laser wavelength. Bottom left panel (b): time evolution of the peak under continuous laser irradiation at $13904 \text{ cm}^{-1}$ (difference in optical depths before and after laser irradiation along y-axis and time in minutes along x-axis). Exponential fit gives a time constant $\approx$ 10 minutes. Bottom right panel (c): corresponding decay of the total luminescence signal (total fluorescence intensity along y-axis and time in minutes along x-axis). Exponential fit gives a time constant of $\approx$ 7 minutes.}
    \label{fig:rexdepdata}
\end{figure}

To better understand the nature of the trapping sites, we experimentally investigated the relaxation process under laser irradiation. For this, we first measured the evolution of the transmission spectrum of a weak white light source. For this, we  irradiate the sample for 8 minutes with a narrow-band continuous-wave laser of \SI{14}{mW} at a fixed wavelength. %The results of three of these measurements are presented in 
Fig. \ref{fig:Relaxation} shows the results of three such measurements with the excitation laser fixed at 12130, 13848, and 11798 $\text{ cm}^{-1}$ that correspond to the given peak resonance positions. 

The measured data show a complex relaxation behavior. A clear depletion is consistently observed at the laser excitation wavelength, along with depletion and growth of bands at other wavelengths. 
If each triplet could be assigned to a specific trapping site, one would expect that pumping one band would lead to the simultaneous depletion of the whole triplet. This is not observed and the interplay between the different features in the spectrum appears to be much more complex. We also observed effects of similar magnitude when the laser was tuned between two bands, and found that the system never fully returned to its initial state, even after hours of white light irradiation. It indicates irradiation effects on the background.
We employed a Gaussian fit with six peaks, sitting upon a variable quadratic polynomial background, to gain a more quantitative analysis during and after the irradiation process. Line intensities, line positions, and linewidths were set as free parameters. %sitting upon a variable quadratic polynomial background was performed during laser irradiation and after it.
The results, for a 20 mW laser irradiation, given in Fig.~\ref{fig:evdecay} indicate 
that the observed fluorescence and transmission peak intensity (under continuous irradiation) show an exponential evolution on a timescale of about ten minutes. But, on the contrary, the s to p excitation time (one over the excitation rate) for a given atom in the matrix remains on the order of 100 microseconds \cite{kanagin2013optical}, since the $\sim 10^6$-fold matrix broadening of the absorption line largely offsets laser intensity that is nearly $10^2$ times greater than the gas-phase saturation intensity of about $1\,$mW/cm$^2$ , leaving the excitation $\sim 10^4$ times slower than spontaneous emission, which is on the order of tens of nanoseconds.
Therefore, the observed evolution on a ten-minute timescale, both with and without laser illumination, clearly indicates that it cannot arise from simple post-excitation dynamics, but it instead reflects much slower, thermally activated processes in the matrix, such as defect or vacancy reorganization, possibly assisted by cumulative heating associated with the Stokes shift between excitation and emission. We therefore investigated further the luminescence properties of the system.

%One possible explanation could be that the Cs atoms are not trapped in well-defined crystalline trapping sites but instead in a more disordered environment. This scenario suggests that some of the observed spectral structures may arise from a distribution of many trapping sites rather than two well-defined environments.
%Indeed,  as shown in the theory part a rich collection of trapping sites is a possible explanation for the relaxation spectra obtained experimentally. 

\section{Luminescence study}

The fluorescence spectra can reveal significant lattice relaxation effects because the initial absorption excites the system into a non-equilibrium state, after which the lattice adjusts prior to fluorescence. 
Differences in the ground and excited state potential energy surfaces result in substantial Stokes shifts \cite{balling1979optical,crepin1999photophysics}. For instance, as shown in Fig. \ref{fig:DIM_Theory_abs}, the DIM theory  predicts a Stokes shift on the order of \SI{3000}{cm^{-1}}, albeit slightly smaller for 6V-O$_{\rm h}$ and slightly larger for 5V-D$_{\rm{3h}}$ trapping sites.

\subsection{Fluorescence  and state relaxation}

\begin{figure}
    \centering
    \includegraphics[width=1\linewidth]{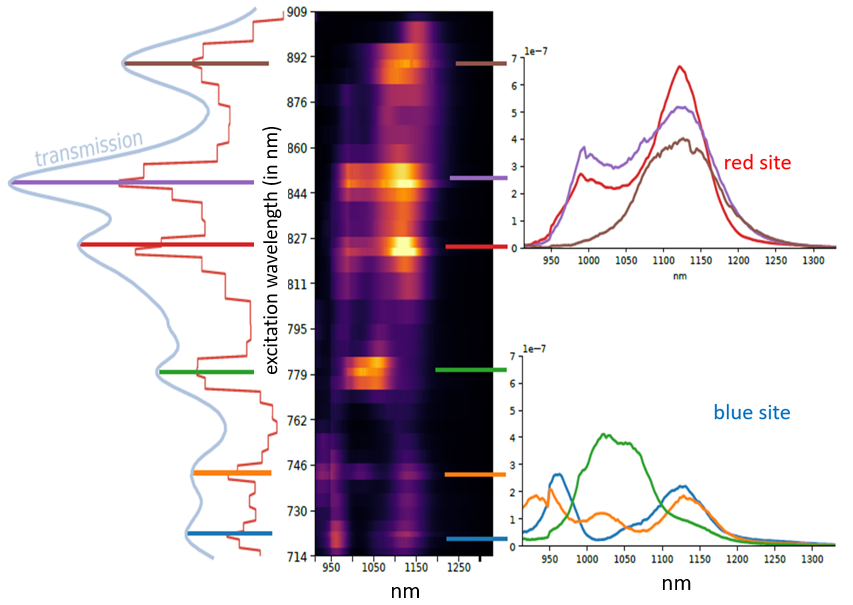}
    \caption{Two-dimensional excitation-emission spectrum of Cs in Ar. The total intensity at each excitation wavelength is plotted on the left side and compared to the transmission spectrum. The fluorescence spectra at each of the six resonance positions are shown on the right side.
    Each emission spectrum has been taken with the same excitation laser power and detector response.}
    \label{fig:2D}
\end{figure}

We performed our measurements using an ANDO AQ-6310C spectrum analyzer, equipped with a grating dedicated to measurements between \SI{400}{nm} and \SI{1750}{nm}.
%It uses one internal grating to measure the intensity of incoming light in a range from \SI{400}{nm} to \SI{1750}{nm}. 
Because of the use of two different diffracted orders, there is a small jump of sensitivity at \SI{560}{nm} and at \SI{950}{nm}. We used the same laser as for the relaxation study and irradiated the sample for 4 minutes at different wavelengths. We observed that the luminescence spectra scaled linearly with the laser power well over the range tested (1 to \SI{100}{mW}). The reproducibility is limited %less obvious, 
but the main features we will discuss remain present
for the different samples. In Fig. \ref{fig:2D} we report a two-dimensional excitation-emission spectrum of Cs in Ar measured at \SI{14}{mW} laser power, for a crystal grown at \SI{8}{K}.

By integrating the luminescence over the entire spectral range and projecting the signal onto the excitation axis, we recover the absorption spectrum (Fig. \ref{fig:2D} left side). This consistency suggests that all excited Cs atoms emit fluorescence and thus no significant non-radiative decay occurs.

We first observe Stokes shift on the order of \SI{3000}{cm^{-1}} very similar to what is predicted by the DIM theory reported in Fig. \ref{fig:DIM_Theory_abs}.
When exciting the red triplet, all lines exhibit fluorescence near \SI{9000}{\per\centi\metre} (\SI{1110}{ nm}). For the $6p_{3/2}$ doublet, this is accompanied by an additional blue-shifted peak (near \SI{1000}{ nm}). 
Exciting the blue triplet leads to even more complex behavior, with multiple emission lines in the  \SI{8800}{}-\SI{11000}{\per\centi\metre}  range, yet the 2 P$_{3/2}$ emission lines retain a distinct similarity compared to those originating from the 2 P$_{1/2}$state.

This multiple emission can have at least three different causes: multiple trapping sites, different relaxation of Cs(6p) internal states or different relaxation of the external degree of freedom matrix. 
The multiple-trapping-site hypothesis is disfavored because each trapping environment is expected to give rise to at least one spin-orbit multiplet. 
Therefore, the observation in Fig. \ref{fig:2D} of only two sets of correlated fluorescence transitions, associated with the two absorption triplets, suggests that the dominant fluorescence arises from two main trapping environments rather than from a larger number of independent sites.
Nevertheless, we note a small background, maybe due to many different low symmetry or grain-boundary trapping
sites, that complicate our measurement and hinder high reproducibility in our samples.

The second hypothesis of
different Cs(6p) internal states modification looks appealing. 
In our case, Fig. \ref{fig:2D}  clearly indicates 
that the $6p_{3/2}$  doublet and $6p_{1/2}$  singlet components show distinct behavior. However, this contradicts several studies of metals in matrix \cite{crepin1999photophysics,ryan2010investigations,davis2018investigation} that show, whatever the initial excitation is, that the electronic population is transferred to the lowest energy adiabatic level within picosecond time scale, usually leading to a single emission band \cite{krylov1994adiabatic,krylov1996nadCl,rojas2003nonadiabatic}. 

To study this effect for Cs in Ar, we performed wave-packet dynamics simulations to analyze the excited-state populations under matrix relaxation following laser excitation. For heavy alkalis such as Cs, the perturbation of spin-orbit (SO) coupling by the Ar environment is weak \cite{battard2023cesium}. We therefore used a DIM electronic structure with position-independent SO coupling, as in previous studies of $^2P$ states in rare-gas matrices \cite{krylov1994adiabatic,krylov1996nadCl}. 
In this context, the model comprises six coupled states of the Cs 6p manifold, treated in a diabatic basis of well-defined angular momentum and spin \cite{krylov1994adiabatic,krylov1996nadCl}, and in which the pair potentials are taken from Refs. \cite{battard2023cesium,jacquet2011spectroscopic,galbis2013potential}. The dynamics are simulated for a Cs atom embedded in Ar matrices (600-800 atoms) beyond twenty picoseconds.

\begin{figure}
    \centering
    \includegraphics[width=1\linewidth]{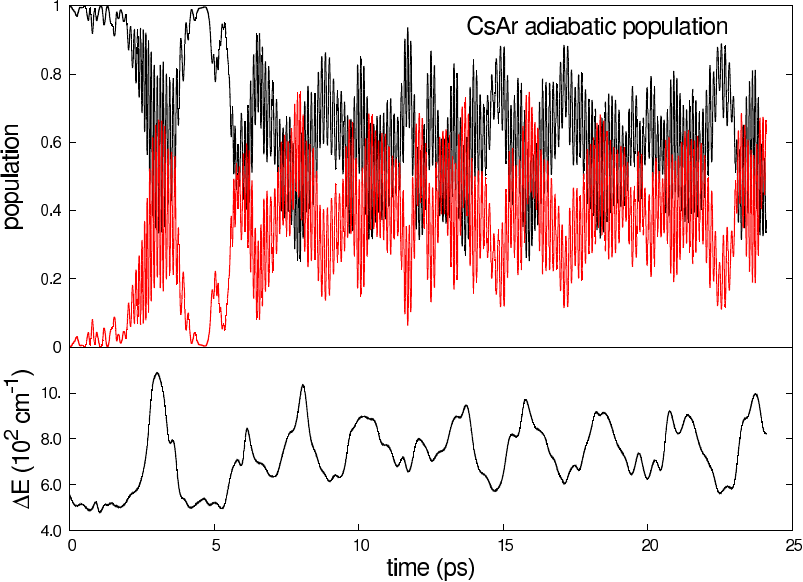}
    \caption{Cs(6p)-Ar adiabatic population evolution right after a Condon (i.e. without nuclear motion) laser excitation from the Cs(6s) equilibrium position. Upper panel: variation of  $J=3/2$ (black: the cumulated population of the 4 uppermost
states) and of $J=1/2$  (red: the 2 lowermost states). Lower Panel: energy difference between these states.}
    \label{fig:dynamics}
\end{figure}

Two trapping sites were considered (T$_{\rm d}$ and O$_{\rm h}$), yielding similar dynamics. Starting from a thermalized 6s ground state at 10 K, the system is promoted to a $J=3/2$ level and propagated for 24 ps. Populations are obtained by projection onto diabatic states.  
The behavior is illustrated in Fig. \ref{fig:dynamics} where we have plotted, together with the eigen-energy difference, the variation of the cumulated population of the 4 uppermost adiabatic states ($J=3/2$) and of the 2 lowermost states ($J=1/2$), for a typical trajectory. Our simulation shows that an initial $J=3/2$ wave packet very quickly acquires $J=1/2$ character (in less than 5 ps). 
Although our adiabatic simulation does not take into account non adiabatic coupling, it indicates time-dependent energy gaps that would transiently enable transitions.  
Because this behavior is similar to what was found in other rare-gas matrix systems,  there is no reason to believe that a full simulation including non-adiabatic coupling  would not lead to the same conclusion that were found in such 
simulations \cite{krylov1996nadCl,rojas2003nonadiabatic,lara2023dynamics}, namely a fast population decay towards the lowest-energy 6p adiabatic state. This excited state will decay by fluorescence towards 6s level producing a single emission band.

% A naive estimation using Landau-Zener formula indicates that the diabatic switching probability $e^{-\pi^2 \nu^2 / (d \nu/dt)}$ in $dt$  at sub-picosecond time scale with a $ \nu \approx \SI{15}{THz}$ energy difference is   negligible and because this 

This is not what is experimentally observed 
and the emissions coming from the two components of the $J=3/2$ doublet point to a more complex luminescence path. We are therefore left only with the last hypothesis of differences in the motion of Cs and Ar particles. 
This has indeed been observed after ${}^1 P_1$ excitation of Hg atoms in  Ar solids \cite{chergui1992spectroscopy} and Zn or Cd in rare-gas matrices~\cite{bracken1997luminescence}, where
the presence of two (or even three) bands in the emission has been
explained by different collective dynamic modes that
are activated in the solid upon photoexcitation of the
metal atom, leading to the coexistence of different energy minima. These minima may be reached by the formation of pseudo-complexes between the metal atom and $n$ rare gas atoms in nearest neighbors positions. 
For instance, in  helium matrix the Cs(6p) atom attracts nearby He atoms, forming Cs$^*$He$_n$ bound states, known as exciplexes that can lead to different emission bands depending on the number $n$ of the bound atoms, and of which formation probabilities depend on the total angular momentum $J$ \cite{moroshkin2006cs,moroshkin2008atomic}.
We may therefore have a similar behavior for Cs in Ar matrices. The different dynamical collective modes in our case may also be correlated with the fine structure during the initial excitation, explaining the difference between the $3/2$ and the $1/2$ excitation. Unfortunately our current theory, which is based on few attempts on different initial conditions (but uses only the D$_{\rm 3h}$ case) in this direction, has not allowed us to identify multiples relaxation points for a given trapping site.

\subsection{Effect of annealing} %experiment}

%A noticeable effect seems that, for blue and red triplet, fluorescence from what we have noted $J=3/2$ doublet (cf Fig. \ref{fig:CsAbs6K}) have quite similar fluorescence behavior; but that itdiffers for the $J=1/2$ one. This may indicate that these peaks do not belong to the same trapping site or that the fluorescence path is not as simple as expected from the theoretical results.\\

In the Zn study, the pair of fluorescence bands observed from single site occupancy presented a temperature-dependence due to the activation barrier of the two paths
\cite{lara2023dynamics}. This example illustrates that changing the sample temperature can be useful for gaining a deeper understanding of the system.

\begin{figure}
    \centering
    \includegraphics[width=1\linewidth]{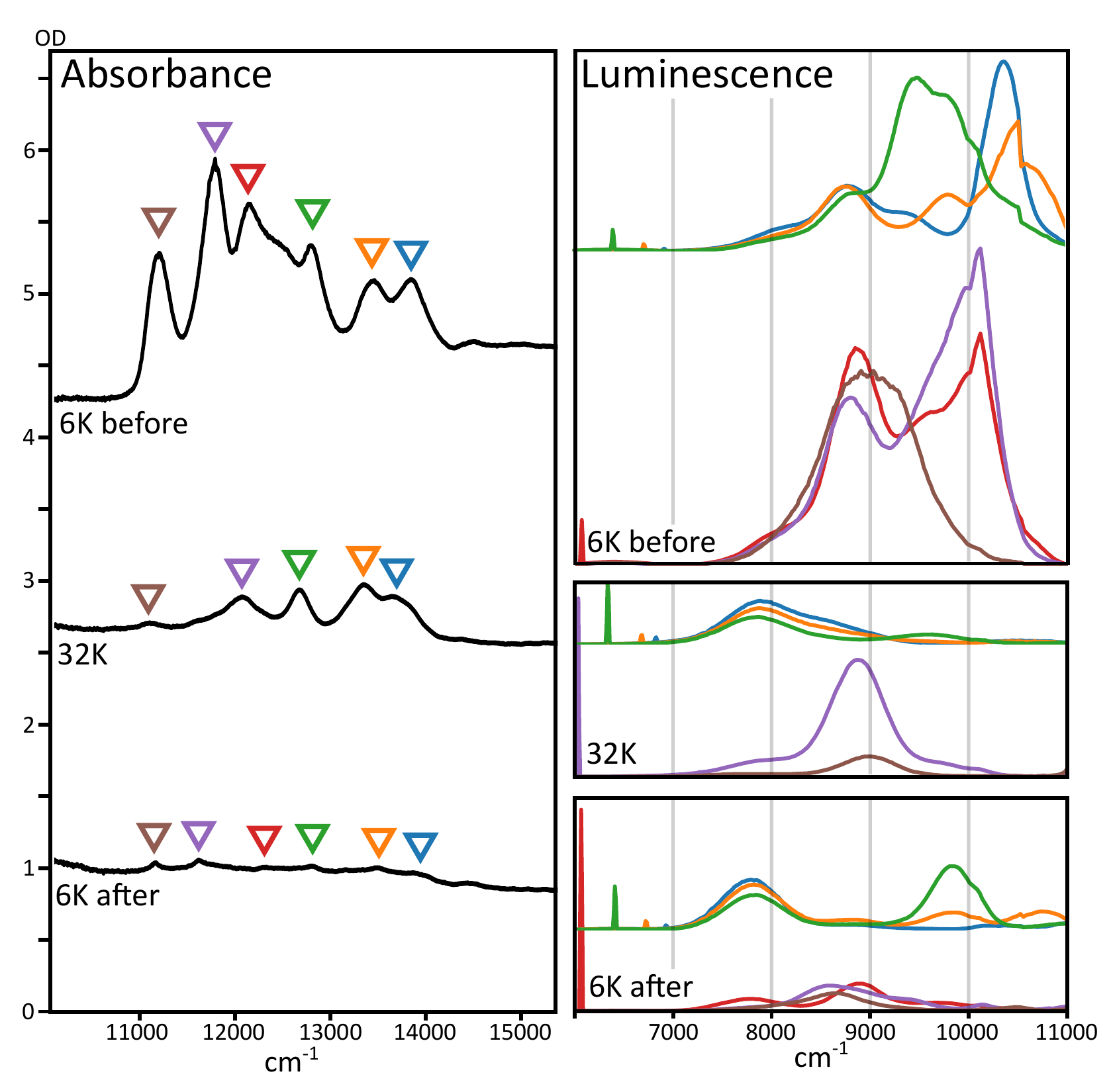}
    \caption{A sample was formed at 6 K, then heated to 32 K, and cooled to 6 K again. The respective absorbance spectra are shown in the left handside (arbitrary offset). At each temperature, the luminescence spectra were taken at the indicated excitation frequencies. They are plotted at the righthand side with a common intensity scale. The sharp peaks at low wavenumbers are artifacts from the laser pump scattering onto the spectrometer grating.}
    \label{fig:Fluo1D_2023_2024}
\end{figure}

We thus performed annealing experiments on an Ar crystal grown at 6 K (see Fig. \ref{fig:Fluo1D_2023_2024}). While the luminescence band positions-which showed no dependence on annealing-remained consistent with previous samples, the relative intensities were not reproducible,  but depended strongly on the alignment of the white light used for illuminating the sample, as well as the sample history (deposition temperature, irradiation history, temperature, etc.). Sample annealing to 32 K led to strong modifications in the transmission spectrum:  The luminescence spectrum simplifies at \SI{32}{K} and shows a single dominant feature for each multiplet.
 This may support the interpretation that the trapping sites undergo an annealing process, leading to a simplified emission spectrum.
Thus, 
while our initial results on Cs in Ar (see \cite{battard2023cesium}) showed little correlation between
the bands assigned to a triplet, the annealed sample gave
 indications that this assignment was correct.  

When cooled back to 6 K, the sample showed strongly reduced absorbance. Although luminescence did not decrease within the same amount, it strongly differed from the initial results, i.e. for the Cs atoms trapped in Ar before annealing. 
This suggests that the Ar environment around the excited Cs atoms participating in luminescence has been strongly modified.

% This also indicates that if the initial, up to now, studied fluorescence arises from two triplets,   some other fluorescence bands exist even if not well-defined features in the absorption are visible, reinforcing the idea of a background of several different low symmetry or grain  boundary types of trapping sites (that are likely to be produced during such fast cooling $32\,$K$\rightarrow 6\,$K process that creates a lot of small crystallites) that absorb light almost in very large wavelength range but emit fluorescence at more defined wavelengths.

%The emission spectra simplify at low Cs densities or higher sample temperatures, suggesting that some spectral features may arise from defect sites or Cs clustering rather than isolated Cs atoms in well-defined environments.

\subsection{Polarization dependence}

To gain further insight into the reorganization of the Cs*(6p)/Ar system in the excited state before emission, we examined the polarization dependence of the emitted light. Indeed, loss of polarization between excitation and emission means a strong reorganization of the guest/host system once excited to a 6p level \cite{andrews2004using}.

\begin{figure}
    \centering
    \includegraphics[width=1\linewidth]{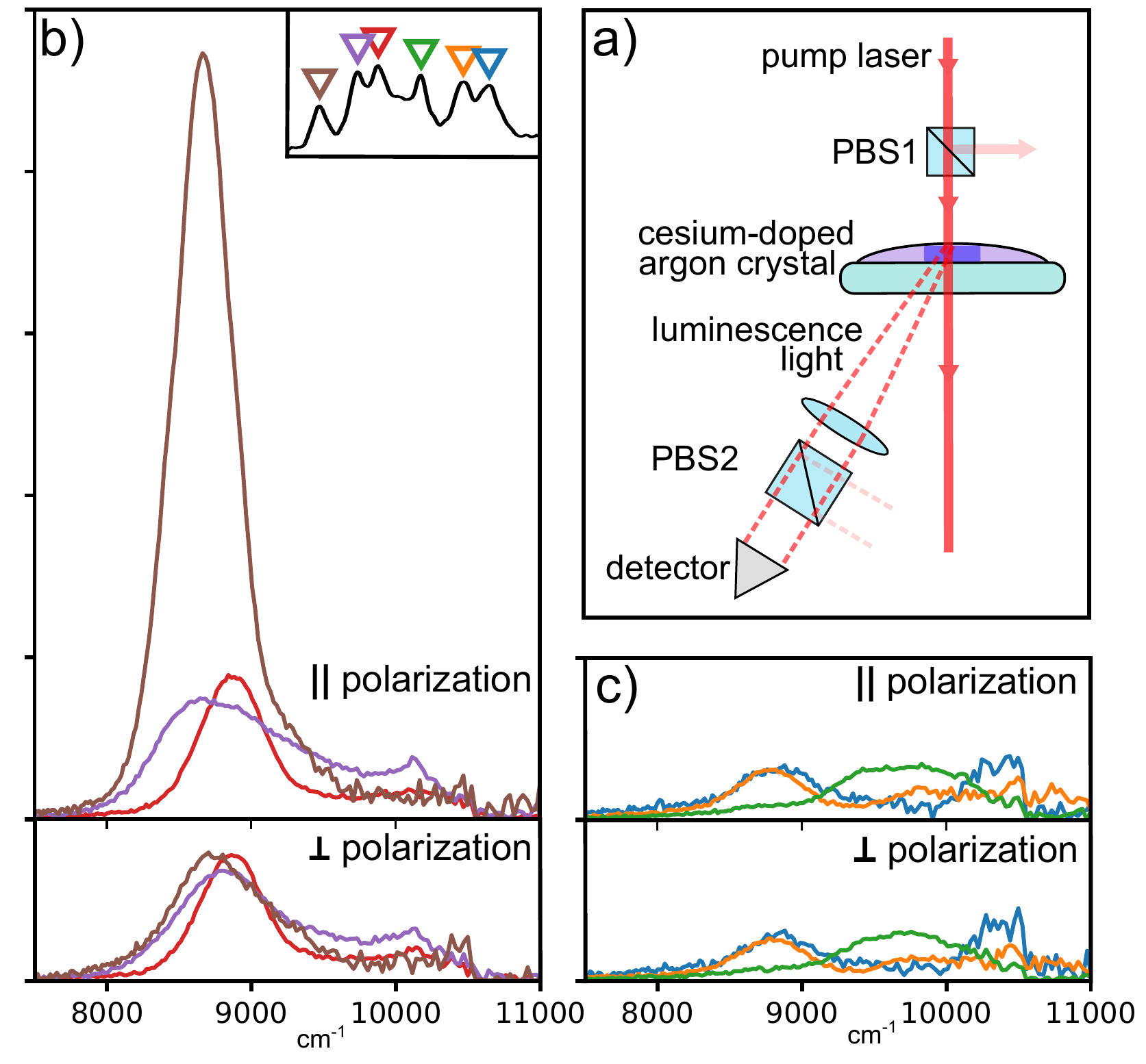}
    \caption{a) Scheme of the polarization-dependent luminescence measurement. We use a Polarizing Beam splitter (PBS) 1 to create a linear polarized pump beam and PBS2 to analyze the emitted light.
    %The luminescence light is guided through PBS2 to enable a polarization dependent detection. PBS2 was rotated into two positions, letting either only light trough that is parallel or orthogonally polarized in reference to the pump.
    b) and c) Luminescence spectra of the six bands with a linear scale for light intensity. There colors correspond to the ones indicated in the inset in the transmission spectrum. }
    \label{fig:polarization}
\end{figure}

%We have thus studied the polarization dependence of the emitted light. 
 The luminescence light was this time guided through a Polarizing Beamsplitter Cube (PBS) acting as a linear polarizer (PBS2 in  Fig. \ref{fig:polarization}). By rotating this PBS by 90 degrees, we were able to select only parallel or orthogonal light with respect  
% only let light pass whose polarization is either linear or orthogonal in reference 
to the pump laser.
 Care was taken when recording such data. Indeed, by looking at the laser excitation light, we first observed that  in our regular thick ($\sim$ \SI{50}{\micro m}) sample, the polarization of the laser was lost due to multiple scattering events. We therefore needed to make a much thinner ($\sim$\SI{10}{\micro m}) crystal where scattered light at the pump laser frequency showed polarization of $\sim 96\%$.

We then measured  the luminescence light emitted while pumping each of the six bands. To account for the smaller luminescence light in the thin samples, we worked at increased laser power. As it differed for each pumping frequency, we normalized the luminescence spectra to the pump power. \\

Fluorescence anisotropy defined as in Ref.~\cite{andrews2004using}, once averaged over the polycrystalline angles, leads to a direct measurement of the angle $\gamma$  between the absorption $\bm{d}_a$  and emission $\bm{d}_e$  transition dipole moments:
\begin{equation}
\mathrm{anisotropy}
= \frac{I_{\parallel}-I_{\perp}}{I_{\parallel}+2I_{\perp}}
= \frac{3\cos^2\gamma-1}{5},
\end{equation}

As shown in Fig. \ref{fig:polarization}, we observed that the luminescence light is completely depolarized except for the $6p_{1/2}$ band farthest into the red. Our results even exceed the  expected maximum of \(I_{\parallel} = 3 I_{\perp}\)  that occurs when the emission and absorption dipole moments are aligned ($\gamma=0$),  possibly hinting at a non-purely averaged polycrystal. However, we are therefore fairly certain, that, for this band, and this band alone, no depolarization occurs between excitation and emission. 

\begin{figure}[t]
  \centering
  \begin{subfigure}[b]{0.49\linewidth}
    \centering
    \includegraphics[width=\linewidth]{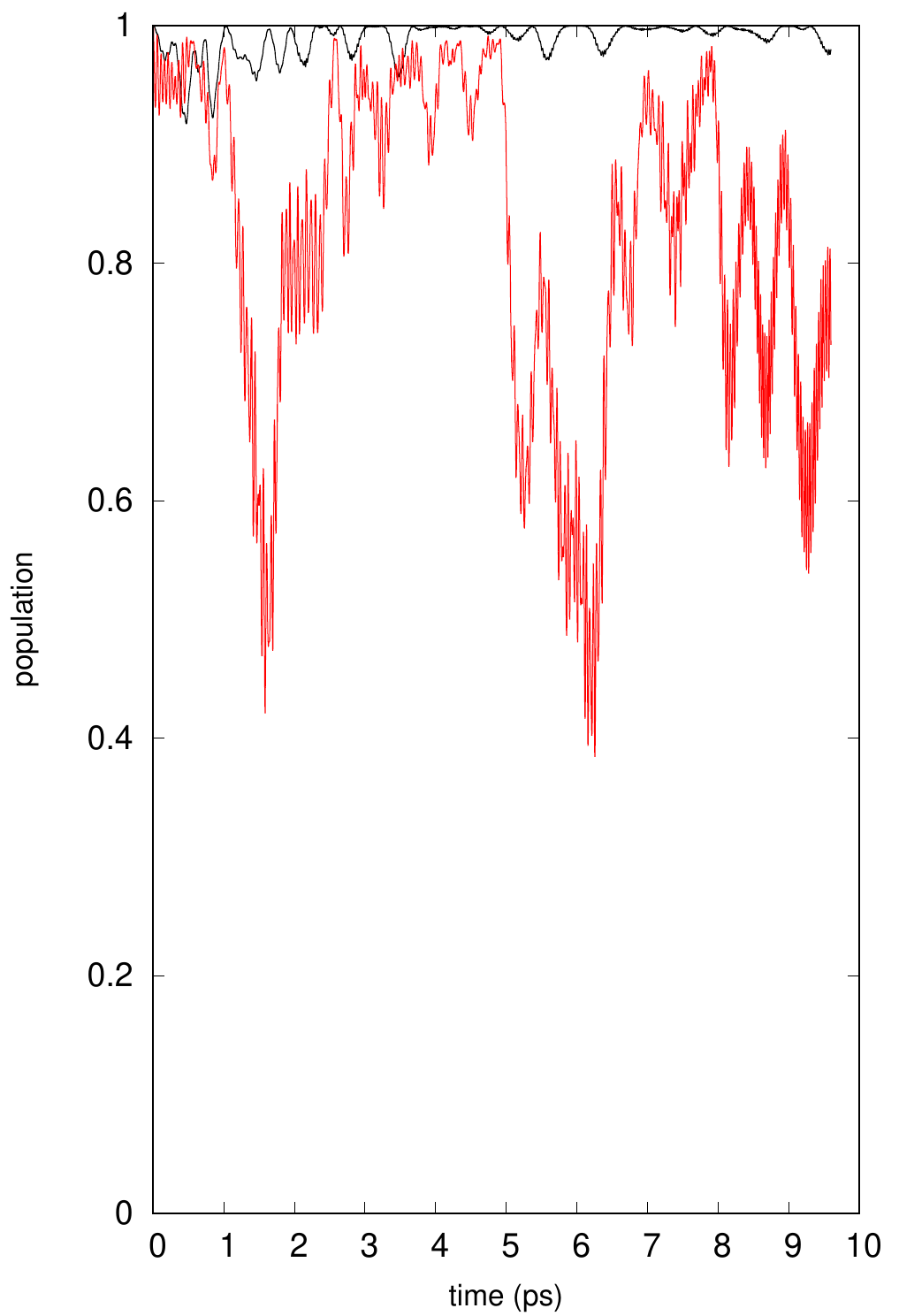}
    \caption{Adiabatic populations (electronic wavepacket decomposition on the eigenstates of the electronic Hamiltonian).}
    \label{fig:adiapopa}
  \end{subfigure}\hfill
  \begin{subfigure}[b]{0.49\linewidth}
    \centering
    \includegraphics[width=\linewidth]{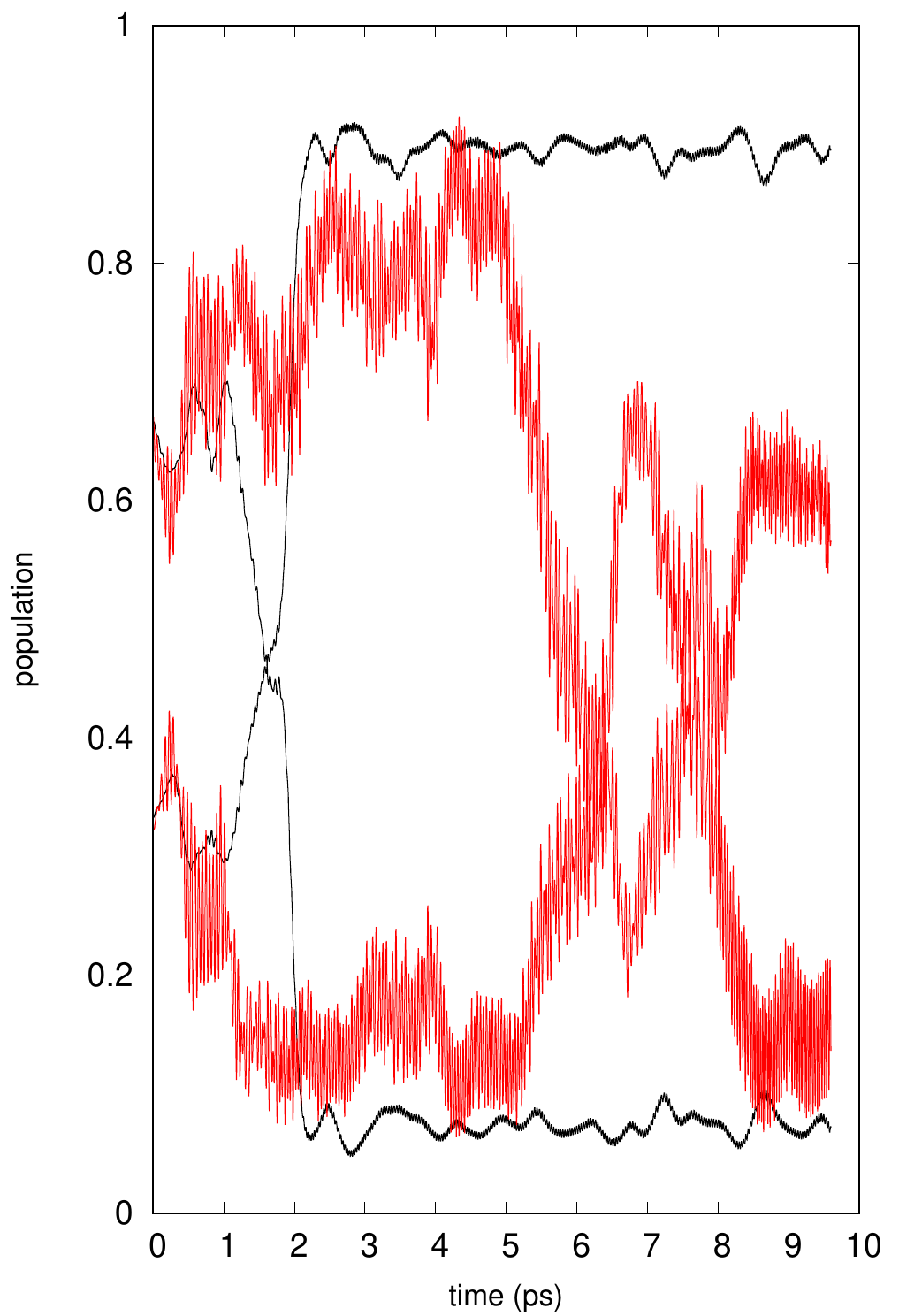} 
    \caption{Projection of the followed adiabatic state onto the uncoupled spin basis \(|M_L,M_S\rangle\).}
    \label{fig:diapop}
  \end{subfigure}
 \caption{Electronic-state population analysis along two independent  trajectories  (black and red lines)
 for the Cs/Ar system in the 6V-\(O_{\rm h}\) site. 
  (a) Adiabatic populations associated with the electronic wavepacket. 
  (b) Decomposition of the followed adiabatic state on the uncoupled basis   \(|M_L,M_S\rangle\): 
  both trajectories start from the expected \(1/3\)-\(2/3\) 
  composition for a 
  \(J=1/2\)
  state.}
  \label{fig:populations_two_traj}
\end{figure}

We may speculate that the red triplet, of which $6p_{1/2}$ component keeps its 6p dipole orientation once excited,  corresponds to a highly symmetric trapping site, such as a cubic $O_{\rm h}$ environment. In such a case, excitation of the spherically symmetric   $6p_{1/2}$ state would induce only minimal structural reorganization, 
thereby preserving the polarization. However, this interpretation is not supported by theory as shown in Figure \ref{fig:populations_two_traj} where  two independent  trajectories (black and red lines) for the Cs/Ar system in the 6V-$O_{\rm h}$  site were simulated. Furthermore,
Figure \ref{fig:DIM_Theory_abs} indicates symmetry change with degeneracy lift in the excited state . 

In Figure \ref{fig:populations_two_traj}(a), the projected population onto  \(J=1/2\) character exhibits (black trajectory)  a weak state mixing, whereas the red trajectory shows substantially stronger mixing, with a time-dependent acquisition of \(J=3/2\) character. 
In (b), the plotted relative weights evolve strongly and differently from one trajectory to the other. These large fluctuations suggest that the dipole orientation is not preserved over time, pointing to a loss of emission polarization. % The latter case is expected in the assumption of the formation of Cs$^*$Ar$_n$ upon excitation.

On the contrary, another possible explanation is that Cs is trapped in a highly constrained site with a well-defined symmetry axis (e.g.\ hcp-, D$_{\rm {3h}}$, grain boundaries-like), such that after excitation, lattice %matrix 
relaxation preserves the dipole orientation while shifting the transition energy. We attempted to confirm this hypothesis theoretically, but found no clear evidence (at least for the D$_{\rm {3h}}$ case). Therefore, we consider this to be a weakly supported hypothesis.

\section{Conclusion}

In this work, we have investigated the fluorescence and relaxation dynamics of Cs atoms embedded in a cryogenic argon matrix using a combination of experimental measurements and DIM-based simulations expanding on our previous investigations \cite{battard2023cesium}. 

Our findings suggest that, although the spectra are superimposed on a complicated background, possibly due to a broad distribution of low-symmetry, defect-related, or grain-boundary trapping sites that might also fluoresce, the main spectral structure is consistent with two dominant trapping environments, producing two triplet features whose doublet (probably mainly of p$_{3/2}$ character) and singlet (probably mainly of p$_{1/2}$ character) components exhibit different fluorescence behavior.
 Polarization studies may indicate that the blue site has
relatively high- or intermediate-symmetry configurations (e.g. T$_{\rm d}$, O$_{\rm h}$) whereas the red  site is possibly of lower-symmetry, such as D$_{\rm {3h}}$ or hcp-like structures (that could explain their relative instability, compared to fcc structure, when heating the sample). However,  a unique assignment remains difficult so further work is clearly needed.

We have confirmed these results using
two different cryogenic instruments in our two laboratories, one equipped with a diamond window and the other with a sapphire window.

In future work, we will employ complementary approaches, including comparison with other rare-gas matrices, electron spin resonance (ESR) spectroscopy, and time-resolved lifetime measurements to further elucidate the nature of trapping environments and the dynamics of excited state relaxation. This will help deepen our  understanding of matrix environment, which is crucial for using the matrix isolation technique as a robust method for probing atomic behavior in controlled environments.

Finally, the observation of persistent fluorescence signals with minimal non-radiative decay, even in complex emission spectra, demonstrates the suitability of these systems for high-precision spectroscopic studies, including searches for fundamental symmetry violations such as the electric dipole moment (EDM).

%Additionally, the observation of persistent fluorescence with negligible non-radiative decay, even in complex emission spectra, demonstrates the exceptional stability and sensitivity of these systems. This makes them particularly well-suited for high-precision spectroscopic investigations, including the study of fundamental symmetry violations, such as the electric dipole moment (EDM).

\section{Acknowledgements}

This research was funded in whole or in part by
l'Agence Nationale de la Recherche (ANR) under the project ANR-21-CE30 -0028-01. 

 A CC-BY public copyright license has been applied by the authors to the present document and will be applied to all subsequent versions up to any Author Accepted Manuscript (AAM) version arising from this submission, in accordance with the grant's open access conditions.

We acknowledge, B. Darquier,  O. Dulieu, H. Lignier, Ch. Malbrunot, B. Viaris for fruitful discussions, and B. Vivan and L. Marriaux for the design and mechanical realization.

%\appendix

%\bibliographystyle{unsrt}
\bibliographystyle{h-physrev}
\bibliography{biblio}

\end{document}